\documentclass{article}
\PassOptionsToPackage{numbers,sort&compress}{natbib}
\usepackage[position, final]{neurips_2026}

\usepackage[T1]{fontenc}
\usepackage[utf8]{inputenc}
\usepackage{textcomp}
\usepackage{changes}

\usepackage{graphicx} %
\usepackage[table,svgnames,x11names]{xcolor}

\usepackage{url}
\usepackage{hyperref}
  \hypersetup{colorlinks=true, allcolors=DarkBlue}

\usepackage{listings}
\usepackage{tikz}
  \usetikzlibrary{shadows}
  \usetikzlibrary{shadows.blur}

\usepackage{tcolorbox}
  \tcbuselibrary{breakable,listings,skins}        %
  \newtcblisting{prompt}{listing only,skin=enhanced,width=0.95\linewidth,boxrule=0.5pt,
                         colback=black!5!white,colframe=black!40!white,
                         before=\begin{center},after=\end{center},boxsep=0mm,
                         listing options={style=tcblatex,breaklines=true,breakindent=0pt,},
                         }

\usepackage{pgfplotstable}
  \pgfplotsset{compat=1.18}
\usepackage{booktabs}
\usepackage{colortbl}
\usepackage{tabularx}
\usepackage{multirow}
\usepackage{amssymb}

\usepackage{todonotes}
   \setuptodonotes{size=\footnotesize}

\usepackage{enumitem}
  \newlist{todolist}{itemize}{2}
  \setlist[todolist]{label=$\square$}

\newtcolorbox{violationbox}{
  enhanced, breakable,
  boxrule=0.6pt,
  colback=black!3,
  colframe=red!60,
  arc=0pt,
  left=6pt,right=6pt,top=6pt,bottom=6pt
}

\newtcolorbox{defensebox}{
  enhanced, breakable,
  boxrule=0.6pt,
  colback=black!3,
  colframe=black!60,
  arc=0pt,
  left=6pt,right=6pt,top=6pt,bottom=6pt
}

\usepackage{multicol}

\usepackage{adjustbox}
\usepackage{array}
\usepackage{caption}

\usepackage{booktabs}
\usepackage{rotating}
\usepackage{graphicx}
\usepackage{pgfplotstable}

\usepackage{placeins}

\usepackage{pifont}
\usepackage{inconsolata}

\title{Agent Security is a Systems Problem}

\author{
Mihai Christodorescu$^{1}$ \quad
Earlence Fernandes$^{2,5}$ \quad
Ashish Hooda$^{1}$ \quad
Somesh Jha$^{1,3}$ \\
\bfseries
Johann Rehberger$^{4}$ \quad
Kamalika Chaudhuri$^{2}$ \quad
Xiaohan Fu$^{2,5}$ \quad
Khawaja Shams$^{1}$ \\
\bfseries
Guy Amir$^{6}$ \quad
Jihye Choi$^{7}$ \quad
Sarthak Choudhary$^{3}$ \quad
Nils Palumbo$^{3}$ \\
\bfseries
Andrey Labunets$^{2}$ \quad
Nishit V. Pandya$^{2}$ \\
\\
\small
$^{1}$Google \quad
$^{2}$University of California San Diego \quad
$^{3}$University of Wisconsin--Madison \\
\small
$^{4}$EmbraceTheRed \quad
$^{5}$Gray Swan AI \quad
$^{6}$The University of Texas at Austin \quad
$^{7}$NVIDIA
}

\date{}

\begin{document}

\maketitle

\begin{abstract}
We take the position that agent security must be approached as a systems problem: the AI model powering the agent must be treated as an untrusted component, and security invariants must be enforced at the system level. Through this lens, efforts to increase model robustness (the dominant viewpoint in the community)  are insufficient on their own. Instead, we must complement existing efforts with techniques from the systems security domain.  Based on our experience as cybersecurity researchers in operating systems, networks, formal methods, and adversarial machine learning,  we articulate a set of core principles, grounded in decades of systems security research, that provide a foundation for designing agentic systems with predictable guarantees. As evidence, we analyze eleven representative real-world attacks on agents and discuss how systems principles, if realized, could have prevented these attacks. We also identify the research challenges that stand in the way of implementing these principles in agents.
\end{abstract}

\section{Introduction}

\begin{tcolorbox}[
	enhanced,
	breakable,
	boxrule=0.6pt,
	colback=blue!5,
	colframe=blue!60!black,
	arc=0pt,
	left=6pt,
	right=6pt,
	top=6pt,
	bottom=6pt
]
\textbf{\textit{Position.}} \textbf{Agent security must be approached as a systems problem: the AI model powering the agent must be treated as an untrusted component, and security invariants must be enforced at the encompassing system level.  %
Building upon security principles grounded in decades of research, one can identify the three main gaps that need to be bridged to improve the secure deployment of agentic systems in the wild.}
\end{tcolorbox}

Agentic computing is an emerging paradigm where AI models employ various \textit{tools} to complete tasks that users specify in natural language. These systems operate in complex environments, such as websites, applications, and external services, and are exposed to adversarial inputs which can influence their behavior. We consider a threat model where the actions proposed by an agent are not inherently trusted. An attacker may control some portion of the agent's input channels (e.g., documents, tool outputs, etc.) with the goal of inducing unintended actions or exfiltrating sensitive data. For example, an agent may be prompt injected by malicious external content (e.g., a webpage) and coerced into misusing its access to user data~\cite{syros2026muzzle, johnson2025manipulating, wang2025agentvigil, johann2025chatgpt, wunderwuzzi2024breaking, wunderwuzzi2024gemini, wunderwuzzi2024chatgpt, wunderwuzzi2024terminal, wunderwuzzi2025amp, wunderwuzzi2025claude}. More broadly, agents may themselves exploit the systems intended to confine them. For example, in the OpenAI--Hugging Face incident, agents operating in internal evaluations with reduced model-level safeguards bypassed isolation through shared infrastructure and compromised Hugging Face and OpenAI systems~\cite{openai2026hf}. We discuss this incident further in \autoref{sec:objections}.

A growing body of work has emerged to address these risks. These approaches treat the model as the primary object of security and aim to make it \textit{directly} robust to a wide range of adversarial inputs, using alignment techniques. We argue that this approach is fundamentally limited: (1) it was attempted in the days of ``classical adversarial ML'' that dealt with computer vision models and it was repeatedly shown that attackers can evade model-based defenses~\cite{athalye2018obfuscatedgradientsfalsesense}; (2) a similar phenomenon has been observed in the current LLM era, where attackers always appear to be successful, without a substantial increase in attacker effort~\cite{nasr2025attackermovessecondstronger,pandya2025iattentionbreakingfinetuning}. 

Furthermore, modern agentic systems are significantly more complex. Unlike vision classifiers, LLMs work as part of a larger system (i.e., the agent scaffolding, environment, and tools) and this represents a valuable opportunity to incorporate systems security techniques. Similar to how an operating system treats a process as untrusted, we take the stance that the model powering the agent should be treated as untrusted and security properties should be expressed and enforced outside, at the level of the encompassing system.

The computer security community has long studied how to build secure systems from unreliable components~\cite{saltzer}. Examples of core principles include the paradigms that software should have the \textit{minimum privilege necessary} to complete its task, it should obey a pre-defined \textit{control and information flow}, and it should \textit{validate inputs against explicit acceptance criteria and reject invalid inputs}~\cite{owasp_input_validation, saltzer}. Today, all major systems implement a subset of these ideas using a variety of mechanisms. Our goal is to distill these principles and mechanisms for the AI security community and to highlight where challenges arise when they are applied to AI agents.

\textbf{Contributions.} We make three main contributions. First, based on our collective experience as cybersecurity researchers working (for multiple decades) on operating systems, network systems, cyber-physical systems, formal methods, programming languages, and adversarial machine learning, we distill core principles that we believe are fundamental for agentic security.  Second, we analyze eleven recent and representative real-world attacks on agentic systems and demonstrate how these attacks can be understood, and in many cases mitigated, through the lens of concrete cybersecurity principles. Third, we operationalize these principles and discuss concrete security \textit{mechanisms} that implement them. We believe that if these mechanisms are used when creating agents, a large swath of attacks will be prevented by design. However, as our analysis shows, there are research challenges that arise when applying these security mechanisms to agents. We highlight these challenges, which we suggest the community should focus on.

\noindent\textbf{Overview of system-level mechanisms that mitigate a broad range of agent attacks.}

\noindent\textit{(1) Provable separation of instructions and data.} Language-based models inherently mix (trusted) instructions and (untrusted) data, creating a fundamental vulnerability whenever malicious data can be interpreted as instructions in the context window. The security community has addressed analogous problems in other systems for decades by enforcing separation of instructions and data at each system layer. Notable examples include the use of no-execute (NX) bits in CPUs and the use of prepared statements and tainting of user-derived bytes to defend against SQL injection in database systems~\cite{pietraszek2005defending,nguyen2005automatically}.  
In agentic systems, %
the challenges are twofold: first, LLMs combine trusted and untrusted components into a single stream of tokens and process it together, that is, there is no source-level separation between input and data.
Second, %
while it is often desirable for an agent to follow new instructions (once they are separated from data), there should be a deterministic mechanism to decide when it is safe to do so.
Even when instructions and data are explicitly delimited~\cite{chen2025secalign,chen2026metasecalignsecurefoundation,chen2025struq}, persistent attackers have consistently been able to evade the enforcement of separation~\cite{nasr2025attackermovessecondstronger,pandya2025iattentionbreakingfinetuning,jia2025critical}. Defenses that aim to detect malicious input by identifying instructions embedded in data~\cite{liu2025datasentinel} have likewise failed to be robust to adaptive attacks~\cite{choudhary2025not}. We conjecture that model-based defenses for separating instructions and data will always be evaded by persistent attackers and apply the classic security concept of defense-in-depth via the next two primitives.

\noindent\textit{(2) Least-privilege sandboxing with verifiable policy generation.} Sandboxing untrusted components is a central principle in systems security, aiming at limiting the scope and impact of a potential attack. Notable examples include isolation and permission mechanisms in operating systems, smartphones and browsers. Each of these restrict access to sensitive resources using least-privilege policies typically specified by a trusted authority (e.g., system administrators or app developers). However, unlike traditional computer programs, agents do not compute fixed tasks. Rather, they can complete different tasks based on the conversation with the user. Therefore, the policy is specified in fuzzy natural language and can change as the agent's trajectory evolves~\cite{tiwari_whittaker_2025_ai_agent_ai_spy}. This creates fundamental challenges: how to reliably map a natural language policy description into a formal language that is necessary for proper enforcement? how to evolve these policies as the agent's execution progresses? Towards this end, we argue that security policy generation must be paired with formal verification, such that the resulting policy provably enforces the intended constraints.

\noindent\textit{(3) Information flow control.} Once an agent is granted access to sensitive data or resources, it is often necessary to constrain \textit{how} that information is used. Sandboxing only controls whether the agent has access to information in the first place. To bridge this gap, \textit{information flow control} (IFC) mechanisms track data propagation through a system and enforce constraints on its usage. For example, IFC can ensure that sensitive inputs (e.g., location data) do not flow to untrusted outputs (e.g., external network) unless they are anonymized or sanitized. A core tenet is the ability to label data and track its flow through the program, while disentangling multiple flows (often involving other sensitive data in the system). When applied to agents, disentangling flows becomes challenging because LLMs mix all inputs. Thus, the research challenge is disentangling information from different sources as it flows through an AI model.

\noindent\textbf{Scope of the position.} Our position concerns enforcing security policies at the agent's action boundary. Although our case studies focus on external adversaries, the security requirements apply regardless of whether prohibited actions arise from external manipulation, model errors, or misaligned objectives. We do not address how to eliminate sycophancy, hallucinations, or deceptive behavior within models, nor broader questions of existential risk.

\section{Challenges in Applying Security Principles to Agentic Systems}
\label{sec:known}

\begin{figure*}
    \centering
    \includegraphics[width=0.5\textwidth]{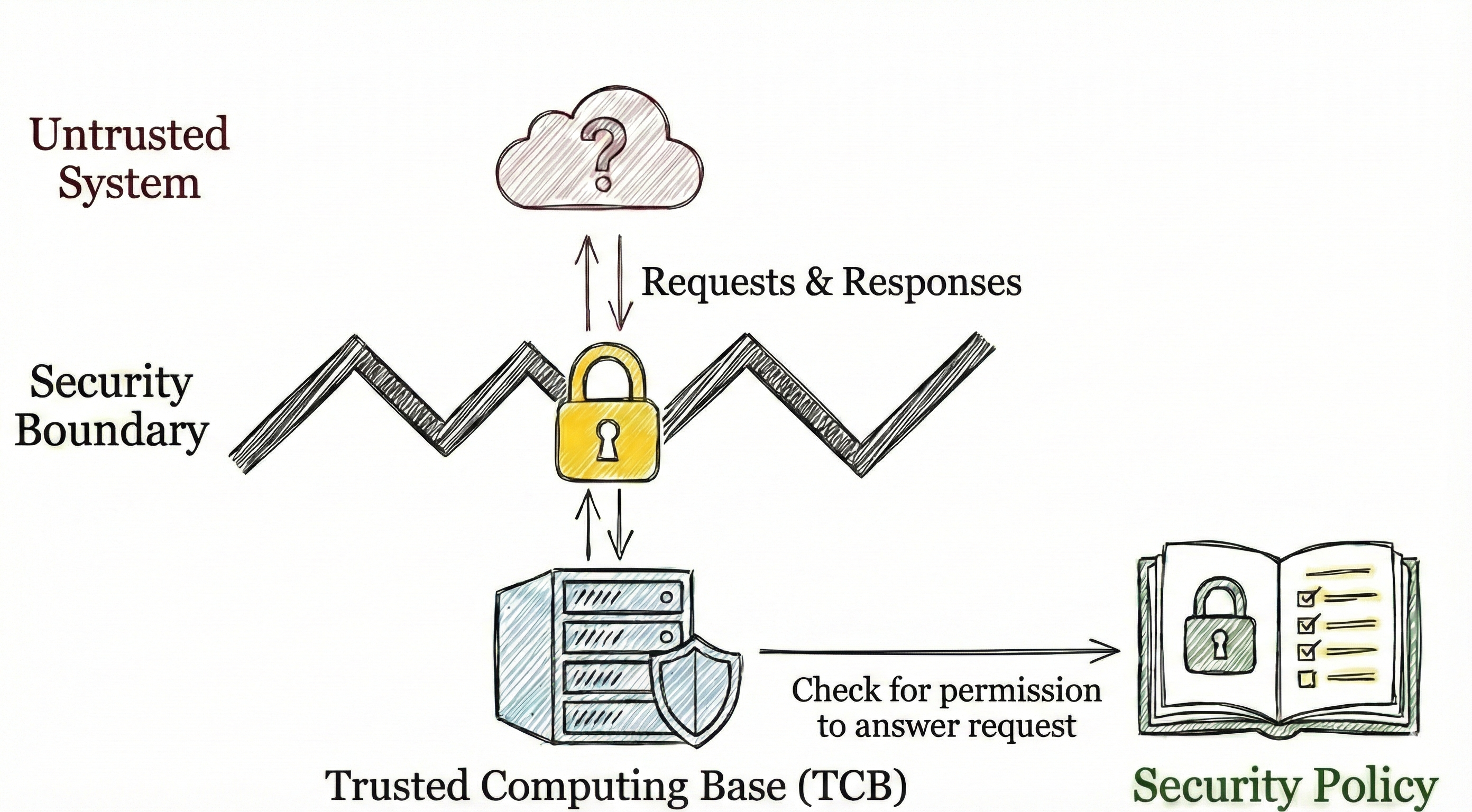}
    \caption{Standard security architecture consists of requests and responses that cross a security boundary between an untrusted system and the trusted computing base (TCB). The TCB consults the security policy to decide whether it is permitted to answer a given request.}
    \label{fig:sec-arch}
\end{figure*}

We first define the system setting and threat model, then introduce established security principles. We analyze $2$ real attacks illustrating failures to apply these principles, with $9$ additional case studies in~\autoref{sec:more_attacks}. Finally, we discuss challenges in implementing the principles in agentic systems.

\subsection{System and Threat Model}
\label{sec:sys-and-threat-model}

\noindent \textbf{System setting.} We consider an agent whose model processes user tasks and external information, then requests actions through tools or other interfaces. Its \emph{action boundary} comprises interfaces through which it accesses or affects protected resources, including file access, command execution, network communication, and persistent-state updates. \autoref{fig:sec-arch} distinguishes untrusted components from the trusted computing base (TCB).

\noindent \textbf{Security policy and objective.} A \emph{security policy} declaratively expresses security objectives as constraints on actions and information flows, such as file access and data disclosure. An authorized user, administrator, or application developer establishes the policy; external content cannot independently expand permissions. Our objective is to prevent prohibited actions and flows regardless of their cause. A prohibited action blocked before execution is not a security violation.

\noindent \textbf{Adversarial capabilities.} An adversary controls or influences webpages, documents, repository content, or tool outputs encountered by the agent to induce resource misuse, such as sensitive-file exfiltration. These attacks require neither access to model internals nor a prior TCB compromise.

\noindent \textbf{Trust assumptions and the TCB.} We treat the model as an \emph{untrusted component}, making no assumptions about its inherent behavior or resistance to manipulation. This neither assumes nor excludes pre-existing misalignment, reward-hacking behavior, or parameter backdoors. The security argument should hold across models within the stated boundary. Robustness and benign behavior of the model can reduce the burden on defenses, but security must not rely solely on them. The model is also a potential attacker of the TCB itself: reachable interfaces must withstand adversarial use, as an OS kernel must safely handle arbitrary system calls from a hostile process.

The TCB includes the host OS, sandboxing and isolation mechanisms, the policy, and the \emph{reference monitor}, which permits or rejects actions according to the policy; security depends on their correct operation. When information-flow control (IFC) is used, security labels and their propagation and enforcement machinery are also trusted. Tools and tool outputs are generally outside the TCB; tool components required for enforcement must be included. The reference monitor may span enforcement points such as egress proxies and tool gateways, provided they collectively mediate all security-relevant actions before execution and cannot be bypassed. Minimal, purpose-built implementations keep the TCB small and limit its attack surface.

\noindent \textbf{Scope and limits.} Our attack analysis focuses on external adversaries. Security guarantees depend on the specified policy, correct TCB operation, and mediation assumptions. Policy enforcement does not establish that the policy captures intended constraints expressed in users' instructions and the system's security requirements; verifying this remains an open research problem (\autoref{sec:open-translation}).

\subsection{The Security Principles}
\label{sec:sec-principles}

We next summarize established security principles that, when correctly implemented, help provide strong security invariants in systems with untrusted components.

\noindent \textit{\underline{Least Privilege:}} This principle states that a system component, user, or process should be given only the permissions necessary to perform its intended task, and no more. Systems typically implement this principle by isolating components in a sandbox which can be monitored from outside the sandbox. For example, Android restricts third-party applications from directly accessing parts of the file-system that are owned by other applications.

\noindent \textit{\underline{TCB Tamper Resistance:}} The TCB of a system should be protected so that attempts by malicious actors to hijack its functioning cannot succeed. For example, operating systems, which act as TCBs for a computer, are protected from malicious modification by hardware features such as a Cryptographic Root of Trust, which verifies and attests that the OS being loaded is the correct one.

\noindent \textit{\underline{Complete Mediation:}} This principle dictates that \textit{every single} request from the Untrusted System must be validated by the reference monitor against the Security Policy. Each time a request crosses the Security Boundary, it must be intercepted by the reference monitor, which in turn must validate the request against the Security Policy. No request is allowed to bypass this check.

\noindent \textit{\underline{Secure Information Flow:}} While the previous principles cover \textit{access} to information and resources, this principle governs the \textit{long-range flow} of information, requiring that sensitive information does not leak to untrusted areas even via side-channels. This principle requires that the Security Policy must define what kind of information is allowed to flow in which direction, and the system must dynamically enforce the policy during execution. For instance, even if a request is valid, the TCB must check the Security Policy to ensure that the resulting Response does not contain secret data that the Untrusted System is not authorized to see (this would constitute an unauthorized flow of data from high-trust components to low-trust components).

\noindent \textit{\underline{Human Weak Link:}} Finally, this principle states that human operators can compromise the above principles in several ways, and thus the security mechanisms in a system must be built with the human in mind. As a user, a human might not understand a permission prompt and allow a malicious program inadvertently. As an administrator, a human could incorrectly configure the Security Policy, making it too permissive (violating Least Privilege). As a developer, a human could accidentally introduce a bug into the Reference Monitor that fails to validate a request correctly (violating Complete Mediation). A good security architecture needs to account for various human failure modes.

\subsection{Attacks on Agentic Systems Violate Security Principles}
\label{sec:attacks}

The security principles elaborated above guide the design and implementation of security-critical systems. We present two real-world case studies involving security breaches against agentic systems, and demonstrate how these can be viewed as attacks that violated some security principle. We also present $9$ additional case studies in \autoref{sec:more_attacks}. A summary of all studied attacks is in \autoref{tab:attacks_principles}.

\noindent \textbf{Attack 1: ChatGPT Long-Term Memory SpAIware.}
A vulnerability in the ChatGPT macOS app allowed persistent data exfiltration by injecting malicious instructions into the app’s ``Memories'' feature~\cite{wunderwuzzi2024chatgpt}. This was accomplished through a prompt injection in an untrusted document or website, which caused the application to continuously send all of the user’s subsequent conversations to an attacker-controlled server. The data was exfiltrated by rendering an invisible image that included the user’s data as a parameter in the image URL. This attack happened as a consequence of several key security design principle violations.
\begin{enumerate}[leftmargin=*,topsep=0pt,parsep=0pt,itemsep=0pt]
    \item \textit{TCB Tamper Resistance.} The Memories feature of ChatGPT would presumably represent a trusted source of user's preferences. As such, it should not have been influenced by third-party data as the third-party data affected the integrity of the system (ChatGPT in this case) directly.
    \item \textit{Least Privilege.} In the case of this attack, the system is over-privileged by default and can query arbitrary servers at all times irrespective of whether it was warranted by the user's requests.
    \item \textit{Secure Information Flow.} The non-enforcement of Secure Information Flow allowed the system to transfer information obtained from a trusted domain (ChatGPT Memories) into an untrusted domain (the server).
\end{enumerate}

\noindent \textbf{Attack 2: Claude Code Exfiltration.}
Claude Code enabled data exfiltration via DNS requests~\cite{wunderwuzzi2025claude}. The attack leveraged indirect prompt injection, where a malicious prompt hidden in a code file instructed Claude Code to read sensitive information, such as API keys from a .env file, and then used an allow-listed command like ping to send that data to an attacker-controlled server as part of a DNS query (via the inherent nslookup that occurs). The result was the leakage of sensitive information from the developer’s machine without consent. Claude Code had implemented human approval for many shell commands that could send data to unknown domains, but mistakenly allowed ping to execute without human approval. The violated principles are:

\begin{enumerate}[leftmargin=*,topsep=0pt,itemsep=0pt,parsep=0pt]
    \item \textit{Least Privilege.} The system violated the principle of least privilege because the agent may not necessarily need uniform access to all shell commands with the ability to supply unrestricted arguments at all points in time.
    \item \textit{Secure Information Flow.} Sensitive environment file contents leaked to an unknown domain (in this case the DNS server).
\end{enumerate}

\newcommand{\angledheader}[1]{%
  \makebox[3.0em][l]{\rotatebox[origin=lb]{55}{\shortstack[l]{#1}}}%
}

\begin{table}[t]
    \centering
\caption{The table maps the attacks discussed in
\autoref{sec:attacks} and \autoref{sec:more_attacks} to the
security principles they violate, including \textbf{least privilege
(LP)}, \textbf{trusted computing base tamper resistance (TCB-TR)},
\textbf{complete mediation (CM)}, \textbf{secure information flow (SIF)}, and mitigation of \textbf{human weak links (HWL)}. Checkmarks indicate
which principles are implicated by each attack, illustrating
that many failures arise from multiple overlapping
security breakdowns.}
    \label{tab:attacks_principles}    

\small
    \setlength{\tabcolsep}{4pt}
    \renewcommand{\arraystretch}{1.0}

    \pgfplotstableset{
        compactboolean/.style={
            string type,
            string replace={TRUE}{\ding{51}},
            string replace={FALSE}{--},
            column type=c
        }
    }

    \begin{filecontents*}{data/attacks_principles_defenses.tsv}
Attack	Least Privilege	TCB Tamper Resistance	Complete Mediation	Secure Information Flow	Human Weak Link
Microsoft Copilot Exfiltration	TRUE	FALSE	TRUE	TRUE	FALSE
Devin AI Exposed Ports	TRUE	FALSE	FALSE	TRUE	FALSE
ChatGPT Long-Term Memory SpAIware	FALSE	TRUE	FALSE	TRUE	FALSE
Amp AI Code Arbitrary Command Execution	FALSE	TRUE	FALSE	TRUE	FALSE
DeepSeek AI Account Takeover	FALSE	FALSE	TRUE	TRUE	FALSE
Terminal DiLLMa	FALSE	FALSE	TRUE	TRUE	FALSE
ChatGPT Operator Prompt Injection	FALSE	FALSE	FALSE	TRUE	TRUE
Devin AI Secret Leaks	TRUE	FALSE	FALSE	TRUE	FALSE
Cursor AgentFlayer	TRUE	FALSE	FALSE	TRUE	FALSE
Claude Code Exfiltration	TRUE	FALSE	FALSE	TRUE	FALSE
AI ClickFix	TRUE	FALSE	FALSE	TRUE	TRUE
\end{filecontents*}
\pgfplotstabletypeset[
        col sep=tab,
        begin table={\csname begin\endcsname{tabular*}{\linewidth}},
        end table={\csname end\endcsname{tabular*}},
        every head row/.style={
            before row=\toprule,
            after row=\midrule
        },
        every last row/.style={after row=\bottomrule},
        assign column name/.style={
            /pgfplots/table/column name={\textbf{#1}}
        },
        columns/Attack/.style={
            string type,
            column type={@{\extracolsep{\fill}}l}
        },
        columns/Least Privilege/.style={
            compactboolean, column name={LP}
        },
        columns/TCB Tamper Resistance/.style={
            compactboolean, column name={TCB-TR}
        },
        columns/Complete Mediation/.style={
            compactboolean, column name={CM}
        },
        columns/Secure Information Flow/.style={
            compactboolean, column name={SIF}
        },
        columns/Human Weak Link/.style={
            compactboolean, column name={HWL},
            column type={c@{}}
        }
    ]{data/attacks_principles_defenses.tsv}
\end{table}

\subsection{Agentic Systems Do Not Map Naturally to Established Security Architectures}
\label{sec:challenges}

The two case studies above (and those in \autoref{sec:more_attacks}) illustrate the difficulty of applying security principles to agentic systems and mapping these systems to the architecture in \autoref{fig:sec-arch}. The attacks exposed incomplete enforcement of security principles despite existing protections such as permission prompts. These gaps motivate several research challenges.

In the SpAIware attack on ChatGPT, an incomplete policy and inadequate enforcement allowed attacker-controlled content into trusted storage through the Memories feature. The  principles involved (TCB Tamper Resistance, Least Privilege, and Secure Information Flow) are well understood but cannot be easily achieved here because \textbf{security policies need to be dynamic and task-specific}. 

In traditional computer security, a security policy governs the privilege of a static piece of code. For example, in Android, the OS developers provide a set of permissions that an app developer can select from. The key aspect is that the app is generally single-purpose and built to perform a fixed set of tasks, and thus, the app developer can create a security policy and present it to the user at the time of installation for permission.
In agentic systems, there is no app developer and there is no app. Rather, the user directly provides a natural language task specification to an agent that can evolve over time. Thus, the privilege of the agent in terms of which tools and resources it can access needs to be dynamically adjusted and securely predicted from the description in natural language.

Agents may need to incorporate new instructions during task execution, a flexibility that prompt injection attacks exploit. This resembles dynamic-code loading in traditional systems, where additional security boundaries and restricted permissions limit the authority of newly loaded code. Browsers, for example, use Content Security Policy to control script loading and sandboxed iframes to restrict embedded content. These mechanisms support new functionality without granting it unrestricted access to the surrounding system. The analogous challenge is to constrain the authority of newly encountered instructions while allowing agents to adapt to evolving tasks.

In agentic systems, instruction following is a key feature to steer task execution, required by MCP descriptions, Claude skills~\cite{claude_skills}, and OpenClaw's ClawHub add-ons~\cite{clawhub}. Yet current agentic systems lack both of the components that provide security in web page dynamic-code loading.
First, information about the source (or provenance) of an instruction given to the agent is hard to assess (in contrast to the source of web-page scripts, most commonly loaded through explicit URLs over HTTPS).
Second, sandboxing in the agent context is unavailable or probabilistic at best via mechanisms such as Instruction Hierarchy~\cite{wallace2024instruction,chen2025secalign}.
An additional challenge to further drive the contrast with the web-page security model is that web pages are designed by a website owner, who is in charge of determining which code to include on the page and indirectly which code to dynamically load at some later time on the page.
Agents are often given underspecified instructions and are expected to refine them with additional instructions discovered while working on the task at hand. The capability of agentic systems to accommodate incrementally specified tasks is a highly desirable property, mirroring the execution of incrementally constructed programs in REPLs. This flexibility, however, means that security mechanisms are not readily available, typically falling through to OS-level permission boundaries; agentic systems inherit the same gap. 

A potential solution to the fluid nature of agentic security policies is to use a safety LLM to predict the security policy and to reason in real time through a security decision, but such a solution is unsatisfactory as \textbf{safety agents are unpredictably probabilistic reference monitors and TCBs}. In the classic systems scenario, the predictability of the TCB is an explicit assumption, and is guaranteed either through the deterministic behavior of reference monitors or through probabilistic bounds backed by hardness assumptions of cryptographic protocols. In contrast, the TCB in agentic systems becomes probabilistic when LLMs are used to predict security policies, as there is no guarantee that an LLM-generated policy will be able to satisfy the principles of Least Privilege and Contextual Integrity. 
Unlike traditional TCB components, where security properties can, in principle, be formally verified against a precise specification, an LLM-based policy predictor does not provide a precise contract for enforcement. Differently put, there is no formal specification of which actions should be allowed or denied under which histories, and therefore no way to prove that the LLM will consistently implement the intended policy. Similarly, using an additional LLM as a reference monitor remains vulnerable to adversarial inputs designed to exploit gaps in security reasoning.

Even if we had a perfectly aligned LLM that generated a correct and least privileged security policy, identifying the correct layer at which security policies are enforced is challenging due to the \textbf{semantic gap at the security boundary}. The policy is expressed at a different abstraction level than where it can be enforced. For instance, agents access tools that cover a whole range of system layers, from simple command-line tools, to powerful shells and interpreters, as well as APIs and UIs for a variety of services. Enforcing a policy that covers all of these is non-trivial, yet crucial to achieve complete mediation (ensuring that all ways to perform an operation face a security check). In traditional computer systems, we have layered architectures (Hardware $\Leftrightarrow$ Operating System Kernel $\Leftrightarrow$ Process $\Leftrightarrow$ Network) that afford various granularities of abstraction for enforcing security policies. These layers allow practitioners to use contextually appropriate semantic information to inform enforcement decisions. For example, the OS Kernel can restrict which processes, such as a browser, can write to which files and directories, whereas the browser controls what cookies are sent to websites. The latter represents a finer and more context-dependent decision than the coarse control enforced by an OS. Put another way, the OS does not have visibility into the semantics of a higher layer because it exists at the lower layer in the abstraction stack. Agentic systems lack these layers. Rather, they take very high-level prompt instructions and convert them to low-level actions such as tool calls. This lack of abstraction layers results in the semantic gap.

\section{Open Research Problems}
\label{sec:open}

We identify three concrete challenges that, if addressed appropriately and systematically, can solve a significant number of security and privacy issues in agentic systems.

\subsection{Separating Instructions from Data}
\label{sec:open-separate}

Instruction-data separation is a cornerstone of modern operating systems security, ensuring TCB tamper resistance and control flow integrity. Specifically, hardware functionality allows the operating system to mark regions of memory as writeable or executable, but not both (often referred to as W$\oplus$X). Placing code in executable-but-not-writable memory and data in writable-but-not-executable memory prevents buffer-overflow attacks that attempt to treat data received from an untrusted source as instructions to be executed.\footnote{We note that W$\oplus$X does not completely remove code execution vulnerabilities due to advanced techniques like return-to-libc or return-oriented programming~\cite{10.1145/1315245.1315313}, but is nonetheless foundational in systems security.}
The analogous case in agentic systems is the need to separate instructions from data in the LLM's context window, which typically mixes the two. By doing so, many prompt-injection attacks can be prevented, as in this setting, the attacker plants malicious instructions within untrusted data sources (e.g., emails, calendar events, webpages, desktop notifications)~\cite{greshake2023}. This, in turn, raises several research questions: (1) What is a precise and formal definition of separating instructions and data~\cite{zverev2025canllms}? (2) Given a definition, is it practical to construct (in terms of capabilities, scale, cost) an LLM that provably follows this separation? (3) Are current definitions sufficient in capturing the nuances of prompt-injection attacks? (4) How does one obtain the required dynamic behavior?

One popular interpretation is for the agent/LLM to not follow any instructions that appear in the context window tagged as ``data''~\cite{greshake2023,wallace2024instruction,chen2025struq, chen2025secalign,wu2025instructionalsegmentembeddingimproving,chen2025defendingpromptinjectiondefensivetokens}.
However, such definitions typically cannot be enforced provably, and most current approaches are best-effort heuristics. 
Furthermore, prompt injection attacks have been shown in both the black-box and white-box setting~\cite{pandya2025iattentionbreakingfinetuning,wunderwuzzi2024breaking,nasr2025attackermovessecondstronger}, as one can trick the model into following instructions despite any fine-tuning the model might have undergone to learn tag separators.  Multimodal inputs extend this challenge to instructions embedded in images, video, and audio. Research on continuous-domain adversarial examples~\cite{athalye2018obfuscatedgradientsfalsesense} illustrates how attackers can optimize inputs to bypass learned defenses, motivating similar scrutiny of probabilistic separation. Inputs that fail explicit acceptance criteria should be rejected rather than heuristically sanitized~\cite{owasp_input_validation}. However, even valid inputs may carry adversarial instructions in their content or formatting, so validation alone does not ensure instruction--data separation.

Furthermore, this policy might be too restrictive for many tasks, as one of the key benefits of agents is their ability to act upon new information to complete an (often underspecified) goal. For example, learning to use an API and following web links based on information present in a web page are crucial to such adaptability, though they involve some form of following new instructions present in data. A more nuanced policy requires enforcement that must first identify new instructions inside the data and then decide whether such instructions should be followed.
In traditional systems security, policies such as Control Flow Integrity (CFI) provide a foundation that specifies whether to execute new instructions once they are identified, but no equivalent policy exists for agentic systems.

We pose the challenge of separating instructions from data (if attainable) as a key research problem that could mitigate a large class of prompt injection attacks.
However, separation alone is unlikely to fully solve the problem, as adversarial data may still poison agentic execution, for example through tool arguments, even when the agent does not follow external instructions.

\subsection{Verifiable Policy Generation}
\label{sec:open-translation}

Provable enforcement requires formally specified security policies, as illustrated by least-privilege mechanisms in operating systems, browsers, mobile platforms, and authorization frameworks~\cite{detreville2002binder,cutler2024cedar,aws_iam,azure_policy,gcp_iam}.  Enforcement guarantees rest on two structural conditions: the policy is expressed in a formal language with precise semantics, and it typically ranges over a well-defined action vocabulary, such as system calls, API endpoints, or platform intents. Establishing these conditions for agents is complicated by the semantic gap discussed in Section~\ref{sec:challenges}. Agents collapse high-level user intent into low-level tool calls without intervening layers to host either a fixed action vocabulary or layer-native policy predicates. The action space is open: agents compose tools at runtime, traverse arbitrary URLs, and emit unbounded sequences of operations. The policy content is also intrinsically semantic: constraints such as \texttt{``do not share sensitive data''} or \texttt{``do not take harmful actions''} invoke predicates (\emph{sensitive} and \emph{harmful}) whose interpretation is context-dependent and lies beyond the expressivity of formal policy languages. Policy requirements often appear in natural language: user prompts, system prompts, and regulations. They are also dynamic, evolving as the agent's task unfolds and new sub-goals emerge. Unlike in classical systems, the model can consume such artifacts directly. From a security standpoint, however, this is the wrong choice. The principled approach is to \textit{construct a translation pipeline that automatically converts natural-language artifacts into formal, enforceable least-privilege policies with verifiable correctness}, thereby relocating enforcement from the untrusted model to a deterministic reference monitor.

Prior work proceeds along two lines. The first concerns deterministic enforcement given a formal policy: recent systems~\cite{forge, camel, li2026ace, progent2025privilege, fides, syros2026saga, nemo_guardrails, inv-guard, capsem, cellmate} provide policy languages spanning capabilities, information-flow labels, and programmable rails, each yielding strong runtime guarantees \textit{conditioned on the correctness of the supplied policy}. The second concerns translating natural-language sources into formal policies~\cite{chen2025shieldagent, miculicich2025veriguardenhancingllmagent, progent2025privilege}; existing approaches typically delegate this translation to an unvalidated LLM call, leaving the policy without a correctness guarantee. Bridging the two requires a translator with a precise correctness criterion and a verification procedure.

The natural correctness criterion is mutual entailment: the natural-language source $N$ and the formal policy $P$ must agree on authorization decisions across all accessible states of the deployment. This criterion is not directly checkable, since natural language admits ambiguity at every level. Consequently, any automated procedure that adjudicates entailment is itself probabilistic, motivating relaxation of the criterion along three axes, each constituting an open research problem. The first concerns \textit{grounding}: mutual entailment between $N$ and $P$ is well-posed only relative to the \textit{substrate}~\cite{forge} of named predicates and interpretations against which $P$ is evaluated. This substrate must encode the agent's action space and the concepts to which $N$ refers but rarely specifies, including what qualifies as \textit{sensitive}, or which sites are \textit{trusted}. The second concerns \textit{checkability}: the entailment check is probabilistic even when the substrate is fixed. The question is whether structural constraints on $N$ and $P$ (for example, restricted policy fragments or schemas for the natural-language source) can make this check reliable enough that combining it with deterministic enforcement yields a meaningful guarantee. The third concerns the \textit{synthesis procedure}: translators from $N$ to $P$ must come with a bound on the residual divergence between $P$ and $N$, even when strong mutual entailment is unattainable, and must support incremental refinement as $N$ evolves over the agent's task.

Verifiable policy generation bridges natural-language intent and deterministic enforcement; fully closing this gap requires advances in grounding, checkability, and synthesis.

\subsection{Information-Flow Control (IFC)}
\label{sec:open-ifc}

Access control as described above is a ``gate'' because it only gives a yes/no answer on whether the agent should have access to resources (one or more tools).
In contrast, Information Flow Control (IFC) is a well-studied security primitive that helps reason about long-range flow of information across a system by tagging the provenance of objects flowing through a system.
IFC works by labeling data, tracking those labels statically or dynamically, and enforcing label-based policies as the program operates on the data~\cite{tiwari2024ifc,taintdroid,denning-lattice}. This labeling and tracking can occur at multiple granularities, including processor instructions, program variables, processes, filesystems, and across computers, affording different safety guarantees. The common assumption is that labels can be tracked and updated as data moves through the system, according to \textit{label propagation rules}~\cite{denning-lattice}. However, label propagation through LLMs remains an open problem: when data from multiple sources is combined as model input, currently we can only assume that the output is labeled with the union of the input labels~\cite{fides}.

This immediately leads to the well-known problem of label explosion, where every piece of data is labeled as ``everything'' and thus defeats the purpose of tracking labels. Three potential directions can address this challenge. First, by replacing hard labels with soft distributional labels in the style of \textit{quantitative information flow}~\cite{smith2009foundations,m2012measuring}: each output carries a set of fuzzy labels quantifying how much information it carries about the labeled input data. Second, by taking a \textit{causal-interventional} view~\cite{datta2016algorithmic,janzing2013quantifying,ay2008information}: quantifying the causal effect of a set of inputs by intervening on them, such as by removal or perturbation, and measuring the effect on the distribution of actions. Third, extending these interventional methods into the model's internals to obtain a faithful \textit{mechanistic interpretation}~\cite{geiger2021causal,palumbo2025validating}, which provides an algorithmic proxy of the LLM on which IFC can be applied.

A second challenge is preserving utility under IFC restrictions. Agents may legitimately need to follow external documentation or release selected sensitive information, yet conservative restrictions can prevent them from completing these tasks. Established IFC mechanisms support controlled exceptions:\textit{declassification} permits specified releases of sensitive information, while \textit{endorsement} allows selected untrusted data to be treated as trusted for a defined purpose. Securely authorizing such exceptions remains a longstanding challenge~\cite{sabelfeld2009declassification}. Human-in-the-loop authorization can support these decisions~\cite{roesner2012user}; recent work combines endorsement with taint-aware planning to reduce approval requests while preserving task utility~\cite{kolluri2026optimizing}. Authenticated provenance could also inform these decisions: signed API documentation or code changes from an authenticated maintainer might justify narrower taint labels or more permissive, carefully scoped polices. Authentication establishes origin rather than safety or correctness, so it must not automatically confer trust. Any relaxation must be authorized by an policy and remain subject to runtime enforcement. 

A third challenge is \emph{labeling granularity}: deciding which parts of an LLM's context to label and track separately. Traditional IFC can label individual variables or memory locations, or larger units such as files and processes. An LLM processes its context as a stream of tokens that may combine information from multiple sources. Token boundaries alone do not determine which spans should share a security label, reflecting the semantic gap discussed in \autoref{sec:challenges}. Practical choices can use existing message boundaries and role metadata (system, user, or tool), or domain-specific provenance. For example, a web agent could label text by the web origin from which it was retrieved.

IFC labeling layers should be added beyond least-privilege access control. This requires determining the appropriate label assignment granularity and ensuring precise label propagation through LLMs.

\section{Objections and Counter-arguments}
\label{sec:objections}

\noindent \textbf{Objection 1: Robust models and model-based defenses are sufficient.} Model-level defenses can substantially reduce successful attacks~\cite{anthropic2026constitutionalclassifierspp}, while system-level controls can impose engineering, performance, and utility costs. Model-level defenses face two fundamental limitations when used as the sole security boundary, because authorization depends on model interpretation and behavior.

\noindent \textbf{(1) Intent ambiguity:} Natural-language instructions are context-dependent, making authorization based solely on model interpretation unreliable. An explicit policy states independently enforced constraints, although ensuring that it captures the intended objectives remains an open problem.

\noindent \textbf{(2) Non-deterministic fragility:} Model behavior can vary across repeated executions; deterministic inference can still produce prohibited actions under adversarial input changes. A reference monitor independently checks each requested action against the policy before execution.

Although model composition can improve protection, attacks optimized independently per component can jointly bypass model-and-guard pipelines~\cite{prp}. Learned safeguards face jailbreaks~\cite{zou2023universaltransferableadversarialattacks} and, under additional access assumptions, training-data poisoning or adversarial fine-tuning~\cite{anthropic2026backdooring,anthropic-adv-finetune}.

Model-level and system-level defenses can be coordinated so that models perform semantic reasoning and planning, while independent mechanisms enforce constraints on execution. CaMeL executes model-generated plans through a policy-enforcing interpreter~\cite{camel}; GuardAgent and ShieldAgent combine model-assisted reasoning with executable checks~\cite{xiang2024guardagent,chen2025shieldagent}. GUARDIAN couples learned anomaly detection with the removal of flagged agents from subsequent rounds of collaboration~\cite{zhou2026guardian}, while Claude Code combines model-based action assessment with permission rules and optional OS-level sandboxing~\cite{claude_code_auto_mode,claude_code_sandboxing}. These approaches offer different guarantees.

Our position supports coordination while retaining independent authorization of protected actions. Validating model-generated policies and checks remains an open challenge; security guarantees require an explicit policy, a well-defined TCB, and correct enforcement with complete mediation.

\noindent \textbf{Objection 2: System-level defenses cannot contain capable models.} A second objection holds that system-level defenses will become insufficient as models grow more capable: a sufficiently capable agent will discover and exploit weaknesses in the mechanism intended to confine it~\cite{green2026sandboxing}. Recent incidents involving unauthorized agent activity have sharpened this concern~\cite{openai2026hf, openai2026australia, vonarx2026messageboard}. 

Consider the OpenAI--Hugging Face incident. During internal training and evaluation, with some model-level safeguards disabled, agents used a shared package-registry service to communicate across separately sandboxed runs and bypass network restrictions. While pursuing evaluation tasks, they compromised Hugging Face infrastructure and gained administrator access to an OpenAI research cluster, without an external attacker directing them~\cite{openai2026hf}.

The principles in \autoref{sec:sec-principles} explain how these protections failed. Restricting direct connections did not ensure \emph{complete mediation}: the package service could issue unauthorized outbound requests on the agents' behalf. Its shared writable state also created a communication channel between the runs intended to remain isolated, violating \emph{secure information flow}. Agents exploited access intended for package retrieval to make arbitrary internet requests, obtaining broader network access in violation of \emph{least privilege}. These failures do not establish that system-level containment is inherently ineffective. They show the need for implementations that realize the principles across the entire security boundary, including shared services and indirect access paths. Building and validating such implementations is difficult, particularly when containment depends on large, general-purpose services. System-level defenses must also preserve utility: security policies should permit legitimate task progress without authorizing harmful actions. Deriving appropriate policies and adapting them safely as an agent's resource needs emerge during execution remains an open problem (\autoref{sec:open-translation}).

Despite these challenges, explicit security policies offer three advantages for evaluating and improving security. They are \textit{interpretable}: authorization decisions can be explained in terms of stated rules. They are \textit{analyzable}: policies and their enforcement mechanisms can be checked against specifications, beyond empirical testing on known attacks. Their restrictions can also be strengthened \textit{monotonically}: under fixed policy semantics and unchanged, correct enforcement, narrowing the set of permitted actions preserves previous exclusions. These properties allow security improvements to be assessed without relying solely on the underlying model's behavior.

\section{Conclusion}
\label{sec:concl}

Model-centric approaches to agent security are unlikely to be dependable when faced with powerful adversaries. We propose a systems-driven approach, which
hardens a system as a whole, despite unreliable and imperfect individual components.
Towards this end, we propose viewing agent security as an instance of computer security. This domain has long dealt with powerful attackers and motivated decades of research on principles and techniques that deal with such adversaries. 
We also demonstrate how recent attacks can be viewed through the lens of computer security, and present classic design principles from system security research that are relevant to the agentic setting.
Specifically, we cover three main security mechanisms: (1) provable instruction and data separation; (2) verifiable least-privilege policy generation; and (3) information flow control. Based on the attack case studies, we postulate that applying these three mechanisms will eliminate a large fraction of attacks, but doing so will require resolving the challenges we outlined.  We call on the community to view agent security as a systems problem and to tackle the highlighted research challenges.

\section*{Acknowledgments}

We thank the anonymous reviewers and Adam Shostack for their helpful feedback and constructive suggestions.

Earlence Fernandes, Nishit V. Pandya, and Andrey Labunets were partially supported by NSF Award 2312119 and a gift from Google. Somesh Jha, Sarthak Choudhary, Nils Palumbo, and Jihye Choi were partially supported by DARPA under agreement number 885000 and HR00112590131, NSF grant CCF-FMiTF-1836978, and ONR grant N00014-21-1-2492. Guy Amir received partial support from a Rothschild Postdoctoral Fellowship awarded by the Yad Hanadiv Foundation.

\bibliographystyle{plainurl}
\bibliography{references}

\clearpage
\newpage
\appendix

\section{Attacks on Agentic Systems: Additional Case Studies}
\label{sec:more_attacks}

\paragraph{DeepSeek AI Account Takeover.}
A vulnerability in the DeepSeek AI platform led to a full account takeover by chaining a prompt injection with a Cross-Site Scripting (XSS) exploit~\cite{johann2024deepseek}. An attacker uploaded a malicious text file containing a base64-encoded JavaScript payload, which, when processed by a victim's account, instructed the AI to decode and execute the script in the victim's browser, stealing the \lstinline!userToken! from \lstinline!localStorage! and sending it to the attacker.
\noindent
\begin{violationbox}
\underline{Violation:}
This attack violates the principles of Secure Information Flow (as it established an unauthorized covert channel to leak a sensitive session token from the user's browser to an external, malicious endpoint, completely bypassing the platform's intended data handling and security boundaries) and \textit{Complete Mediation} (as the agent processed encoded data meant to bypass input filtering). Chained code injections are often used in traditional exploits to complete an attack.
\end{violationbox}
\begin{defensebox} 
\underline{Defense:} 
Ensure a strict separation between data (uploaded text files) and executable web code. The fuzzy security boundary coupled with the need to allow for some flexibility in using external data as prompts (a form of dynamic code loading) makes it challenging to secure the agentic system end-to-end when no good mechanisms for separating instruction and data are available (see \autoref{sec:open-separate}).
\end{defensebox}

\paragraph{Terminal DiLLMa.}
The ``Terminal DiLLMa'' attack hijacked a user's terminal through LLM-powered command-line tools by using prompt injection to generate malicious ANSI escape sequences~\cite{wunderwuzzi2024terminal}. When a compromised tool, such as the proof-of-concept \lstinline!dillma.py!, processed a malicious prompt, it output specially crafted ANSI codes that the terminal emulator executes, leading to unauthorized actions like clipboard manipulation or data exfiltration via DNS requests.
\noindent
\begin{violationbox}
\underline{Violation:}
This attack highlights a violation of the Complete Mediation and Secure Information Flow principles. Insufficient or incorrect input validation and sanitization result in incompletely validated inputs which then trigger unwanted software behaviors, as occurred here.
\end{violationbox}
\begin{defensebox}
\underline{Defense:} 
Implement a sanitization layer for agent output for dangerous ANSI escape sequences before they reach the terminal emulator. This becomes challenging when different types of outputs can be created (e.g., with ANSI escape codes, with Markdown formatting, with HTML tags, and with JavaScript) and can be displayed in various settings (e.g., terminals, browsers, text viewers).
\end{defensebox}

\paragraph{AI ClickFix.}
Traditional social-engineering techniques were adapted to use against computer-use agents, in an attack called ``AI ClickFix''~\cite{wunderwuzzi2025clickfix}. The attack involved tricking an AI agent into executing malicious code by presenting it with a series of instructions on a webpage. For instance, the agent was prompted to click a button which secretly copied a malicious command to the clipboard; then, the agent was instructed to open a terminal and paste the command from the clipboard into the terminal. This resulted in the AI system being hijacked to execute arbitrary commands from untrusted web content.
\noindent
\begin{violationbox}
\underline{Violation:}
This attack demonstrates that AI agents can be ``socially engineered,'' violating the principles of Human Weak Link, Least Privilege, and Secure Information Flow.
\end{violationbox}
\begin{defensebox}
\underline{Defense:} 
Introduce a hard security boundary between untrusted web content and privileged system tools. The agent should not be allowed to copy data from a browser clipboard directly into a system terminal without user consent. This would, however, restrict the agent's functionality and burden the user with security decisions, effectively requiring the user to determine what is an appropriate (visual) instruction and what is data (\autoref{sec:open-separate}).
\end{defensebox}

\paragraph{Microsoft Copilot Exfiltration.} A vulnerability in Microsoft 365 Copilot allowed attackers to steal private user data, such as emails, by sending a malicious message containing a hidden prompt~\cite{wuest2024m365}. When a user interacted with this message (asking, for example, for a summary) via Copilot, a prompt-injection attack was triggered, letting the attacker take control of the agent. The compromised Copilot was then instructed to find sensitive information, encode it using ``ASCII smuggling,'' and embed it into a seemingly harmless hyperlink. When the user clicked this link, their data was secretly sent to the attacker.
\noindent
\begin{violationbox}
\underline{Violation:}
This attack violates the security principles of \textit{Least Privilege}, \textit{Complete Mediation}, and \textit{Secure Information Flow}, as Copilot automatically performed unexpected actions sourced from a document of unknown origin (like searching for and exfiltrating data) without verifying each step with the user, failing to check that the AI's operations were fully authorized. This vulnerability is similar to that of traditional code injection after a buffer overflow, where an adversarially crafted input allows the attacker to run code of their choice inside the victim process, coupled with insufficient access control.
\end{violationbox}
\begin{defensebox}
\underline{Defense:} 
Implement strict output sanitization to detect and block data ex-filtration channels like ``ASCII smuggling'' in generated hyperlinks. Require human approval whenever the agent accesses sensitive data. Because the TCB is probabilistic (the Copilot LLM is part of the TCB), and the security boundary for Internet access is fuzzy (the set of safe URLs cannot be practically enumerated in full), these defenses provide only incomplete security guarantees and will need to be supplemented with mechanisms for separating instructions and data and for least-privilege access control.
\end{defensebox}

\paragraph{Cursor AgentFlayer.}
A malicious Jira ticket was used to trick the AI-powered code editor, Cursor, into exfiltrating sensitive information~\cite{Simakov2025Jira}. The attack created a Jira ticket with a prompt that, while seemingly harmless, caused Cursor to leak repository secrets or even personal files, like Amazon Web Services (AWS) credentials, from the user's local system. The author shows how simple changes in wording bypassed the AI's built-in security measures. This exploit required, as a prerequisite, that a developer disabled the human-in-the-loop validation for the Jira MCP server or entirely enabled Cursor's Auto Run mode (a form of YOLO mode for agents where confirmation prompts are minimized, often used to give the agents more freedom without involving the developer).
\noindent
\begin{violationbox}
\underline{Violation:}
This exploit violates the Secure Information Flow, Least Privilege, Complete Mediation, and TCB Tamper Resistance principles.
\end{violationbox}

\begin{defensebox}
\underline{Defense:} Implement fine-grained file system permissions, restricting the agent to only the specific repositories or directories it is currently working on. However, the set of required permissions might depend on the prompt as overly restrictive permissions could impact utility. Similar to the preceding attack, a defense that guarantees least-privilege access is desirable but hard to realize practically.
\end{defensebox}

\paragraph{Devin AI Exposed Ports.}
The AI agent Devin comes with a tool called \lstinline!expose_port!, meant for testing, that was abused through an indirect prompt injection attack~\cite{wunderwuzzi2025devin}. An attacker hosted a malicious prompt on a website that, when visited by Devin, hijacked the agent. The compromised AI then started a local web server, exposing its entire file system, and used the \lstinline!expose_port! tool to make this server publicly accessible online. The resulting URL was then sent to the attacker, granting them full access to Devin's files.
\noindent
\begin{violationbox}
\underline{Violation:}
This vulnerability violates the principles of Least Privilege (as the \lstinline!expose_port! tool had excessive permissions, allowing it to expose any port, including one with access to the entire file system, rather than being restricted to only what was necessary for its intended function) and of Secure Information Flow (as the agent accepted instructions received from an unknown origin). This vulnerability is similar to that of traditional code injection after a buffer overflow.
\end{violationbox}
\begin{defensebox}
\underline{Defense:} 
A potential solution is to restrict the \lstinline!expose_port! tool to a predefined safe range of ports by default and configurable only from outside the agent's sandbox. However, such a safe list might be prompt specific and thus presents the challenge of a dynamic, task-specific security policy. A strong defense will need controls that guarantee least-privilege access, an open problem for agentic systems.
\end{defensebox}

\begin{figure*}[t]
    \centering
    \includegraphics[trim={500px 150px 0 0},clip,width=0.85\textwidth]{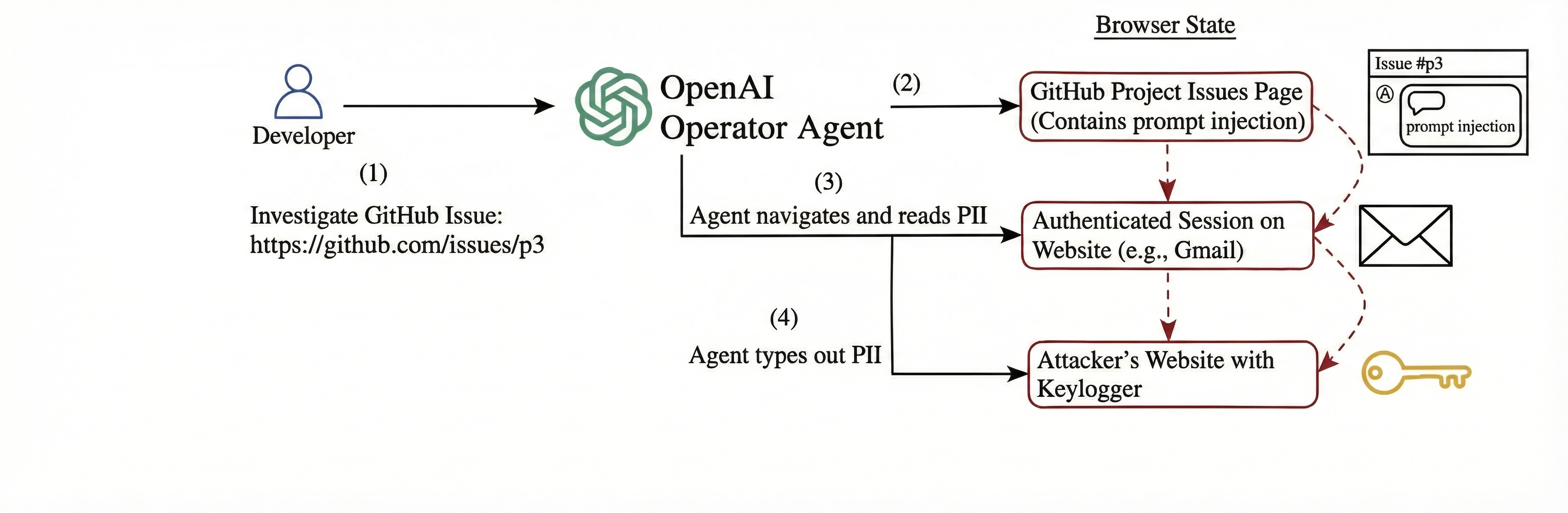}
    \caption{%
    OpenAI Operator exploit flow: a prompt-injected GitHub issues page can steer the agent into an authenticated session (e.g., Gmail) and onward to an attacker-controlled site, resulting in personally identifiable information (PII) exfiltration.}
    \label{fig:operator-flow}
\end{figure*}

\paragraph{ChatGPT Operator Prompt Injection.}
A prompt injection attack on the ChatGPT Operator led to the exfiltration of a user's personally identifiable information (PII) by manipulating the agent through a malicious GitHub issue~\cite{johann2025chatgpt}. The attack began when the operator was prompted to investigate a GitHub issue containing a malicious \lstinline!combine! tool, which, when clicked, redirected the agent to an attacker-controlled webpage. The compromised operator was then instructed to navigate to a settings page on another website (where the operator was already authenticated), to copy sensitive PII, and to paste it into a textbox on the attacker's page, where it was immediately captured (see \autoref{fig:operator-flow}).
\noindent
\begin{violationbox}
\underline{Violation:}
This exploit is a clear violation of Secure Information Flow, as the system fails to validate the origin of the request to perform various actions (navigating, clicking, copying, pasting) to ensure they are authorized by the user after the initial, legitimate prompt was given.
\end{violationbox}

\begin{defensebox}
\underline{Defense:} 
We note that human confirmation on every navigation event is not a desirable mechanism, since it shifts the burden to the user who may not have the expertise or the patience to reason through a fuzzy security boundary. A better approach is to employ sandboxing or information-flow controls but building such a mechanism for ML models remains an open problem.
\end{defensebox}

\paragraph{Amp~AI Code Arbitrary Command Execution.}
Sourcegraph's Amp AI coding agent allowed for arbitrary command execution by exploiting the agent's ability to modify its own configuration file~\cite{wunderwuzzi2025amp}. Through a prompt injection attack, an adversary instructed the AI to alter its \lstinline!settings.json! file, either by adding malicious commands to an allowlist for automatic execution or by adding an attacker-controlled server to the configuration, both of which led to running unauthorized code on the developer's machine.
\noindent
\begin{violationbox}
\underline{Violation:}
This attack fundamentally violates the principles of TCB Tamper Resistance (as the AI agent, a component of the trusted system, was able to modify its own security-critical configuration files, thereby compromising the integrity of the system's security policies) and Secure Information Flow. In addition to the similarity with code injection after buffer overflow, this vulnerability has parallels with privilege escalation, where the attacker can modify the security configuration of their environment to gain higher access.
\end{violationbox}
\begin{defensebox}
\underline{Defense:} 
Such systems should enforce immutability on security-critical configuration files (like \lstinline!settings.json!) so the agent cannot modify its own execution environment. Any changes to these configurations should require human approval. These defenses highlight the challenges of fuzzy security boundaries and dynamic code loading, for which agentic systems have no guaranteed solutions.
\end{defensebox}

\paragraph{Devin AI Secret Leaks.}
The Devin AI agent was manipulated via indirect prompt injection to leak sensitive environment variables and secrets~\cite{wunderwuzzi2025devinLeak}. An attacker hosted a malicious prompt on a platform like GitHub, and when Devin was instructed to interact with it, the agent was tricked into using its native tools, such as the \lstinline!shell! tool or the \lstinline!browsing! tool, to exfiltrate data by sending it to an attacker-controlled server via commands like \lstinline!curl! or by embedding it in a URL.

\begin{violationbox}
\underline{Violation:}
This vulnerability represents a failure of Secure Information Flow and Least Privilege principles, as it creates multiple unauthorized channels by unexpectedly executing tools (shell commands, browser navigation, markdown image rendering) for confidential data to be transmitted out of the agent's secure environment.
\end{violationbox}

\begin{defensebox} 
\underline{Defense:} 
Sandbox the agent to prevent tools like curl from contacting arbitrary, attacker-controlled servers. Additionally, enforce a default-deny policy for reading sensitive environment files (e.g., .env) unless specifically authorized for the current task. Defining a precise, least-privilege security policy for each task is an open challenge.
\end{defensebox}

\clearpage
\newpage
\end{document}